\documentclass[11pt,a4paper]{article}
\usepackage{jcappub}
\usepackage{booktabs}

\title{Improved Sensitivity of the DRIFT-IId Directional Dark Matter Experiment using Machine Learning}

\author[a]{J.B.R. Battat,}
\author[b]{C. Eldridge,}
\author[b]{A.C. Ezeribe,}
\author[a]{O.P. Gaunt,}
\author[c]{J.-L. Gauvreau,}
\author[b]{R.R. Marcelo Gregorio,}
\author[a]{E.K.K. Habich,}
\author[a]{K.E. Hall,}
\author[d]{J.L. Harton,}
\author[a]{I. Ingabire,}
\author[e]{R. Lafler,}
\author[e]{D. Loomba,}
\author[b,1]{W.A. Lynch$^{\ast}$,}
\author[f]{S.M. Paling,}
\author[a]{A.Y. Pan,}
\author[b]{A. Scarff,}
\author[d]{F.G. Schuckman II,}
\author[c]{D.P. Snowden-Ifft,}
\author[b]{N.J.C. Spooner,}
\author[f]{C. Toth}
\author[a]{and A.A. Xu}

\affiliation[a]{Department of Physics, Wellesley College, 106 Central Street, Wellesley, MA 02481, U.S.A.}
\affiliation[b]{Department of Physics and Astronomy, University of Sheffield, Hounsfield Road, Sheffield, S3 7RH, U.K.}
\affiliation[c]{Department of Physics, Occidental College, 1600 Campus Road, Los Angeles, CA 90041, U.S.A.}
\affiliation[d]{Department of Physics, Colorado State University, Fort Collins, CO 80523-1875, U.S.A.}
\affiliation[e]{Department of Physics and Astronomy, University of New Mexico, 800 Yale Boulevard, Albuquerque, NM 87131, U.S.A.}
\affiliation[f]{STFC Boulby Underground Laboratory, Boulby mine, Loftus Saltburn-by-the-sea, Cleveland, TS13 4UZ, U.K.}
\affiliation[1]{Current Address: Department of Physics, University of York, Heslington, York, YO10 5DD, U.K.}

\emailAdd{warren.lynch@york.ac.uk}

\abstract{We demonstrate a new type of analysis for the DRIFT-IId directional dark matter detector using a machine learning algorithm called a Random Forest Classifier. The analysis labels events as signal or background based on a series of selection parameters, rather than solely applying hard cuts. The analysis efficiency is shown to be comparable to our previous result at high energy but with increased efficiency at lower energies. This leads to a projected sensitivity enhancement of one order of magnitude below a WIMP mass of 15 GeV c$^{-2}$ and a projected sensitivity limit that reaches down to a WIMP mass of 9 GeV c$^{-2}$, which is a first for a directionally sensitive dark matter detector.}


\notoc
\begin{document}
\maketitle
\flushbottom

\section{Introduction}

A considerable amount of evidence suggests that $\sim$84\% of the total mass content of the Universe is accounted for by dark matter \cite{doi:10.1142/S0218271817300129}. A favoured hypothesis is that this matter is comprised of so-called Weakly Interacting Massive Particles (WIMPs) \cite{doi:10.1146/annurev-astro-082708-101659}. DRIFT-IId (Directional Recoil Identification from Tracks) is an experiment searching for low energy recoils caused by WIMP-nucleus interactions. However, unlike most detectors, DRIFT-IId is sensitive to the direction of nuclear recoil events induced by elastic scattering of WIMPs \cite{ALNER2005173}. These recoils can then be compared to the expected WIMP-wind direction \cite{PhysRevD.37.1353}, providing a means to unambiguously detect a dark matter signal. Previous analysis of DRIFT data has been used to establish the detector's sensitivity \cite{BATTAT201765}. The analysis presented here leverages machine learning techniques to reduce the amount of signal events (mimicked using a neutron source) lost to data reduction, therefore improving the detector's sensitivity while preserving DRIFT's excellent background rejection {\cite{BATTAT201765}}. This type of analysis will become essential for larger and more costly experiments (such as that proposed by the CYGNUS collaboration \cite{Vahsen:2020pzb}), which will most likely incorporate a more complex readout configuration \cite{EZERIBE2021}.   

\section{The DRIFT-IId Detector \label{sec:detector}}
 
\begin{figure}[t]
	\centering
	\includegraphics[width=\textwidth]{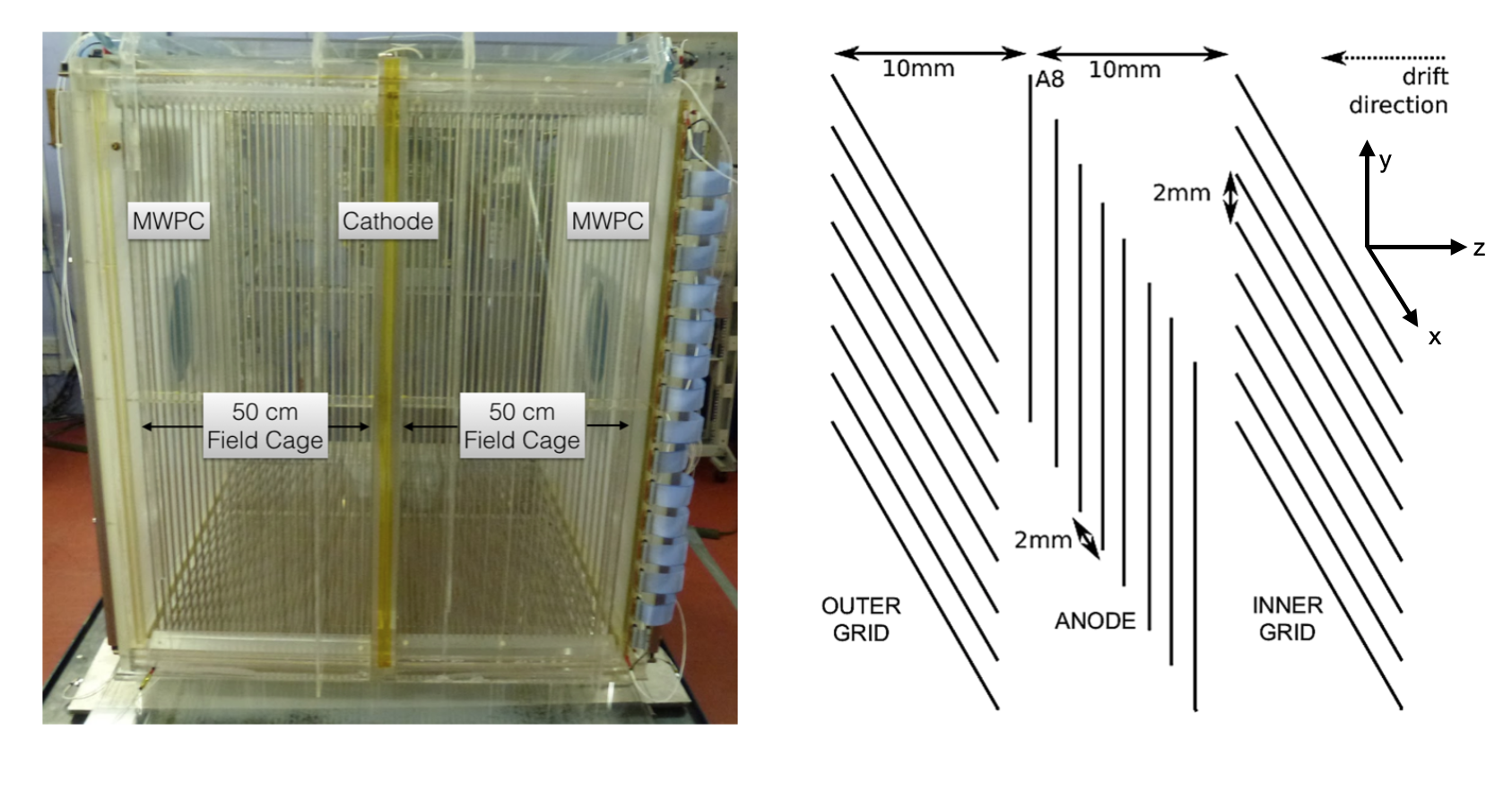}
	\caption{Left: DRIFT-IId NI-TPC. Each MWPC, left and right, is separated from a central cathode via a 50 cm field cage. Right: Schematic showing part of one of the two MWPCs used by DRIFT-IId. Made from three arrays of 552 stainless steel wires of 100 $\mu$m (grid) and 20 $\mu$m (anode) diameter. The wire pitch of each array is 2 mm and the separation between the arrays is 1 cm.}
	\label{fig:DRIFT+MWPC}
\end{figure}

DRIFT-IId is a 1 m$^{3}$ NI-TPC (Negative-Ion Time Projection Chamber) located at the Boulby Underground Laboratory. A detailed discussion of the DRIFT apparatus can be found in refs. \cite{BATTAT201765, ALNER2005173, DAW2012397, BATTAT20151}. Briefly, the detector, shown in Figure \ref{fig:DRIFT+MWPC} (left), consists of two MWPC (Multi-Wire Proportional Chamber) readouts placed 50 cm away from, and either side of, a central cathode. The figure (left) also shows the field cage used to smoothly reduce the voltage between the cathode and MWPCs, creating a uniform drift field of 580 V cm$^{-1}$. Each MWPC is composed of three stainless steel wire arrays of 2~mm pitch, including an anode array of 20~$\mu$m thick wires, and two grid arrays of 100~$\mu$m thick wires placed orthogonal to the anode. The configuration of the arrays is shown in Figure \ref{fig:DRIFT+MWPC} (right). The grid and anode wires are separated by a 1~cm gap and are held at -2.884 kV and ground, respectively. This produces a high electric field of up to 3~kV~cm$^{-1}$ within the gap which is used to produce signal amplification via electron avalanche. The signal is then collected by the anode wires, which in turn induces a response on the inner grid wires (see Figure \ref{fig:DRIFT+MWPC} (right)). The measured signal on the anode and grid wires then provides track information in the x, y and z directions.

For each array, 52 wires on the outer edges are used to veto events entering or exiting the fiducial volume of the detector and to guard against electrical breakdown at the array extremities. For the signal wires, every 8th wire is grouped in order to minimise the amount of processing electronics. The grouping size was chosen as neutron calibrations showed that no recoil within the energy region of interest ($<$200 keV$_{r}$) is expected to trigger more than 8 wires. Each group of signal and veto wires is processed by a Cremat-110 pre-amplifier and Cremat-200 (4 $\mu$s) shaper, before being recorded by the Data AcQuisition system (DAQ).

\begin{figure}[t]
	\centering
	\includegraphics[scale=0.47]{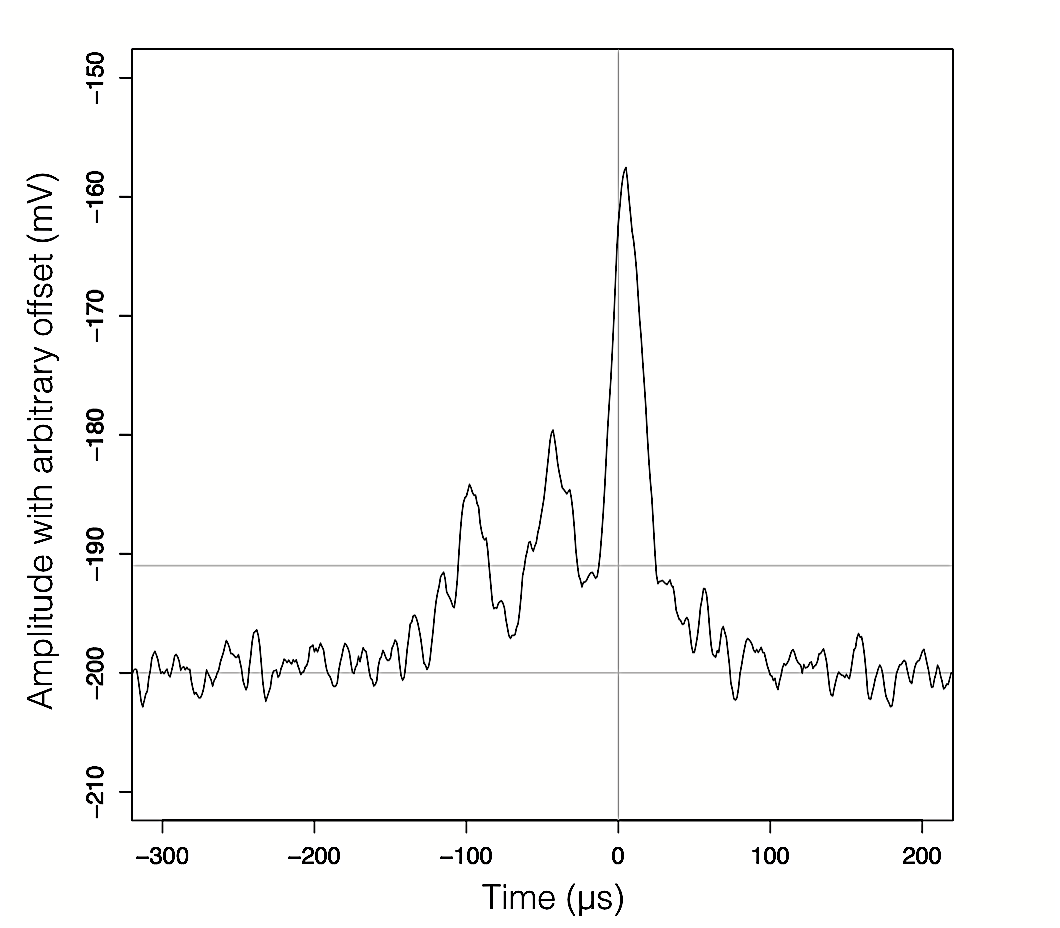}
	\caption{A typical neutron-induced recoil event, showing a main peak and two smaller minority peaks, as seen on a single anode wire.}
	\label{fig:mp}
\end{figure}

The DRIFT-IId detector is located inside a 7 mm thick steel vacuum vessel \cite{ALNER2005173}. The vessel is surrounded by polypropylene pellets, which provide shielding from neutrons produced during the radioactive decay of isotopes in the surrounding rock walls. Gamma shielding is not used. Instead, a combination of signal threshold and short shaping time prevents the low ionisation density of Compton scattered electrons from triggering a response. This allows for a gamma rejection of 1.98$\times{10^{-7}}$ with a threshold of 1000 NIPs (Number of Ionised Pairs) \cite{BATTAT201765}. The vacuum vessel is evacuated and back-filled to a pressure of 41 Torr, utilising a gas mixture of CS$_{2}$, CF$_{4}$ and O$_{2}$ with partial pressures of 30, 10 and 1 Torr, respectively. The CS$_{2}$ provides negative ion drift \cite{MARTOFF2000355}, CF$_{4}$ provides a spin-dependent (SD) fluorine target, and the addition of O$_{2}$ enables fiducilization along the drift direction, as described by ref. \cite{doi:10.1063/1.4861908}. Briefly, the latter is achieved due to the presence of minority peaks within the signal waveform, such as those shown in Figure \ref{fig:mp}. These peaks are produced by the creation of unique anion species, caused by the addition of O$_{2}$, which have slightly different drift velocities and therefore arrive at the MWPC at different times. The separation between the peaks is used to calculate the recoil's position along the drift direction, which combined with the planar information from the MWPC readout, allows for fiducialization in 3D.

A common background source for dark matter detectors are Radon Progeny Recoils (RPRs), which result from the decay of $^{222}$Rn gas. This background source and its mitigation has been covered extensively in previous publications by the DRIFT collaboration, such as ref. \cite{BATTAT201533}. In summary, the implementation of a 0.9 $\mu$m thick aluminised-mylar cathode, along with fiducialization cuts to tag and remove events occurring within 2 cm of the cathode, allows for the rejection of RPR events. The ability to veto backgrounds in three dimensions ensures that the target volume of gas is fully fiducialized. Further details can be found in refs. \cite{BATTAT20151} and \cite{BRACK2015130}. 
 
\section{Data Selection and Calibration \label{sec:DRIFT_data_select}}

Supervised Machine Learning (ML) works by training and testing an algorithm on data that is known to be, in this case, either signal or background (see Section \ref{sec:DRIFT_ML}). To simulate signal data, DRIFT-IId was exposed to a $^{252}$Cf neutron source \cite{BURGOS2009417}, placed 10 cm above, and at the centre of the top surface of the TPC vessel. The source produced neutrons at a rate of $2.8\pm{0.2}\times{10}^{3}$ s$^{-1}$ \cite{BATTAT201765}, a portion of which entered the fiducial volume and caused nuclear recoils that mimicked a WIMP signal. A total of 0.9 days of neutron exposure was used to train and test the algorithm on signal recognition. For the background data, 155 days of DRIFT-IId WIMP search data was used, all of which was previously analysed and shown to produce no WIMP signal candidate and therefore only include background events \cite{FREDTHESIS}. As this background data was produced without using a radioactive source, it is referred to here as source free. Along with this data, background obtained during three days of exposure to three $^{60}$Co sources, placed on top of the vessel was also included. Table \ref{tab:data_select} lists the recoil data discussed and gives the usage as either background or signal as well as the total live time in days and the total number of recorded events. 

\begin{table}[t]
	\begin{centering}
	\caption{DRIFT-IId runs used for the analysis.}
	\label{tab:data_select}
	\begin{tabular}{llrr}
	\toprule
	Run & Usage & Days & Events\\
	\midrule
	Neutron & Signal & 0.90 & 51550 \\
	Source free & Background & 155 & 597172 \\
	Co-60 & Background & 2.98 & 34112 \\
	\bottomrule
	\end{tabular}
	\par\end{centering}
\end{table}
\vspace{0.5cm}

The avalanche field produced by the grid and anode arrays of the MWPCs (see Section \ref{sec:detector}) causes multiplication of each ionisation electron (gas gain) by a factor of $\sim$1000 \cite{Battat_2014}. The DRIFT electronics then amplifies this signal further. To calculate the original NIPs value produced by the recoil event, the detector was calibrated by regularly exposing the fiducial volume to two $^{55}$Fe sources, located behind each MWPC. The sources were placed behind an automated shutter that opened every six hours for approximately three minutes. During this exposure the 15 mV hardware threshold was removed to enable 5.9 keV electron recoils, caused by the photoabsorption of $^{55}$Fe X-rays, to be recorded. This produced a signal of known energy that was converted to NIPs using the gas mixture W value (25.2~$\pm$~0.6 eV \cite{PUSHKIN2009569}) and compared to the recorded pulse integrals. 

\section{Recoil Discrimination Parameters \label{subsec:DRIFT_params}}

After some initial waveform processing, involving smoothing and the removal of high and low frequency noise (originating, respectively, from the cathode and mains supply), any event passing a hardware threshold of 15 mV was recorded by the DAQ. This section describes the reconstructed event parameters derived from the recorded waveforms that were used to classify events as either signal or background. \par

Before the ML stage of the analysis, an initial data reduction stage (stage 0 cuts) was conducted to remove events that could be described by any of the following: triggered one or more of the veto wires, and so, originated outside of the fiducial volume; triggered 8 or more wires, which corresponds to an ionisation trail of $\geq$16 mm, a WIMP induced recoil is only expected to produce an ionisation trail of a few mm; triggered both sides of the detector simultaneously, which is unlikely for a WIMP event; produced non-contiguous wire hits, a nuclear recoil is expected to produce a contiguous response; produced $>$6000 NIPs, which corresponds to a WIMP velocity that exceeds the galactic escape velocity; had a calculated drift distance of $\leq$11 cm (see Section \ref{sec:detector}), which is too close to the readouts to accurately resolve the minority peaks (see Figure \ref{fig:mp}); occurred within 2 cm of the cathode and could, therefore, be an RPR event (see Section \ref{sec:detector}). As one of the main aims of this ML analysis was to extend the WIMP search capability to low mass ($<$ 10 GeV c$^{-2}$), no lower energy threshold was implemented other than the hardware threshold mentioned above. \par

This initial stage of data reduction reduced the fiducial region along the drift direction from 0-50~cm to 11-48~cm. Along with the veto wire region around the MWPCs, this created a DRIFT-IId fiducial volume of 0.59~m$^{3}$ and a total fiducialized fluorine mass $M_{SD}$ of 24.1 g. After the stage 0 cuts were applied the ML algorithm was used to classify the remaining events as either signal or background. The reconstructed event parameters (features) used for the ML analysis possess a range of different values (as opposed to the boolean values used in the stage 0 cuts). For example, the parameter describing the number of triggered anode wires has a value of between 1 and 7 (after the stage 0 cuts). These parameters, which are listed in Table \ref{tab:cont_cuts}, were used as input for the ML analysis and were, therefore, labelled ML parameters features.

\begin{table}[t]
	\begin{centering}
	\caption{The ML parameters used to classify events in the DRIFT-IId data set.}
	\label{tab:cont_cuts}
	\resizebox{0.75\columnwidth}{!}{
	\begin{tabular}{l|p{0.7\linewidth}}
	\toprule
	Parameter Name & Parameter Description \\
	\toprule
	Anode NIPs & The ionisation energy collected on the anode wires, in NIPs. \\
	\midrule
	d & The distance from the MWPC to the ionisation event vertex. \\
	\midrule
	Max Pulse Height & The maximum pulse height on the anode wires. \\
	\midrule
	Pulse Width & The width of the anode pulse with the maximum pulse height (Full Width Half Maximum).\\
	\midrule 
	Pulse Area & The integrated area of the anode pulse with the maximum pulse height.\\
	\midrule
	Anode Hits & The number of anode wires with signal above threshold. \\
	\midrule
	Risetime & The time duration between 10\% and 90\% of the maximum pulse height recorded on the anode wires. \\
	\midrule
	Peak Ratio & The ratio between the minority peak integral and the main peak integral (see Figure \ref{fig:mp}) for the anode wire with the maximum pulse amplitude. \\
	\midrule
	Grid NIPs & The response induced on the grid wires, converted to NIPs using the calibration method described in Section \ref{sec:DRIFT_data_select}. \\
	\bottomrule
	\end{tabular}
	}
	\par\end{centering}
\end{table}
\vspace{0.5 cm}

\begin{figure}[t]
	\centering
	\includegraphics[width=\textwidth]{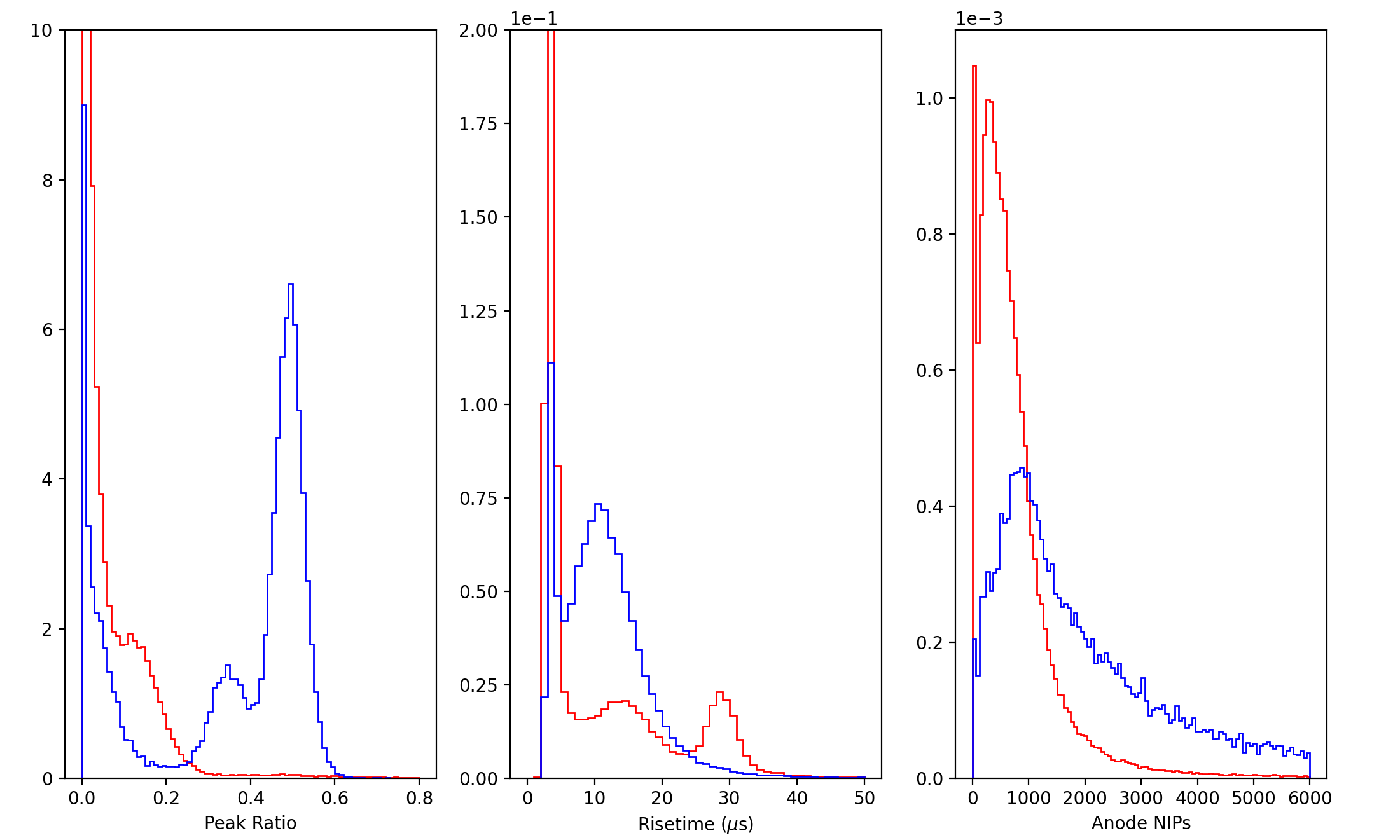}
	\caption{Probability density histograms for three of the parameters listed in Table \ref{tab:cont_cuts} that show the best background (red) to signal (blue) discrimination after the stage 0 cuts were applied.}
	\label{fig:param_hists}
\end{figure}

Figure \ref{fig:param_hists} shows probability density histograms for three of the ML parameters listed in Table \ref{tab:cont_cuts} that showed the best background (red) to signal (blue) discrimination. The ML algorithm leverages the differences in the signal and background distributions for each parameter to tag/label unclassified events from DRIFT.

The standard way of producing cuts for the parameters listed in Table \ref{tab:cont_cuts} would be to investigate the best cut positions for each individual parameter like those shown in Figure {\ref{fig:param_hists}}. For the analysis described in the next section, the ML algorithm treats all parameters collectively to produce a more efficient background rejection model.

\section{RFC Analysis Algorithm \label{sec:DRIFT_ML}}

\begin{figure}[t]
	\centering
	\includegraphics[width=\textwidth]{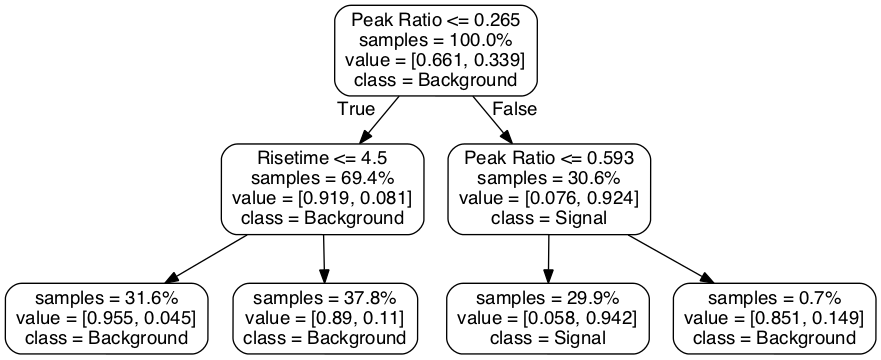}
	\caption{The first couple of data reduction levels produced by one of the DT's used in the analysis. The parameter cut is shown at the top of each box, `sample' shows the percentage of events remaining, `value' gives the proportion of background (right) and signal (left) events and `class' shows the majority event type.}
	\label{fig:tree}
\end{figure}

A machine learning algorithm, called a Random Forest Classifier (RFC) \cite{MACHLEARN}, was used. The algorithm is based on the Decision Tree (DT) \cite{MACHLEARN} method for finding the best parameter cut positions that maximise signal to background separation. The type of DTs used for this analysis employ the Gini Index method for decision making {\cite{5994250}}. This method decides which parameter and parameter value to use to split the data such that the split data purity is maximised. For example, the DT shown in Figure {\ref{fig:tree}} first splits the data using the Peak Ratio parameter with a parameter value of 0.265. As shown in this figure, the majority of events remaining after producing a true response to this selection criteria are background events, however some events that produce a true response are signal. By including the possibility that these events can be later classified as signal by further selection criteria, the DT algorithm may still correctly label an event as signal even though it would have been tagged as background by a standard analysis.

\vspace{0.5 cm}

The RFC algorithm extends the DT algorithm by producing multiple DTs using the signal and background recoil parameter data described in Section \ref{sec:DRIFT_data_select}. It then computes an averaged result from all of the trees to provide a better overall classification scheme, compared to that of a single tree. The accuracy of the analysis can be optimised by setting two ML hyperparameters: the tree depth and number of DTs. The tree depth selects the number of DT levels used by the analysis, for example Figure \ref{fig:tree} shows a two-level DT. Selecting a depth too small would limit the decision tree's ability to separate signal from background, whilst selecting a depth too large would overfit the data during training and produce a less accurate result when applied to new data. Increasing the number of trees used by the RFC creates a more accurate averaged result. However, this also increases the CPU time involved and, at some point, a larger number of trees either no longer improves the result or provides such a small improvement that the trade-off in CPU time is not beneficial. 

\begin{figure}[t]
	\centering
	\includegraphics[scale=0.7]{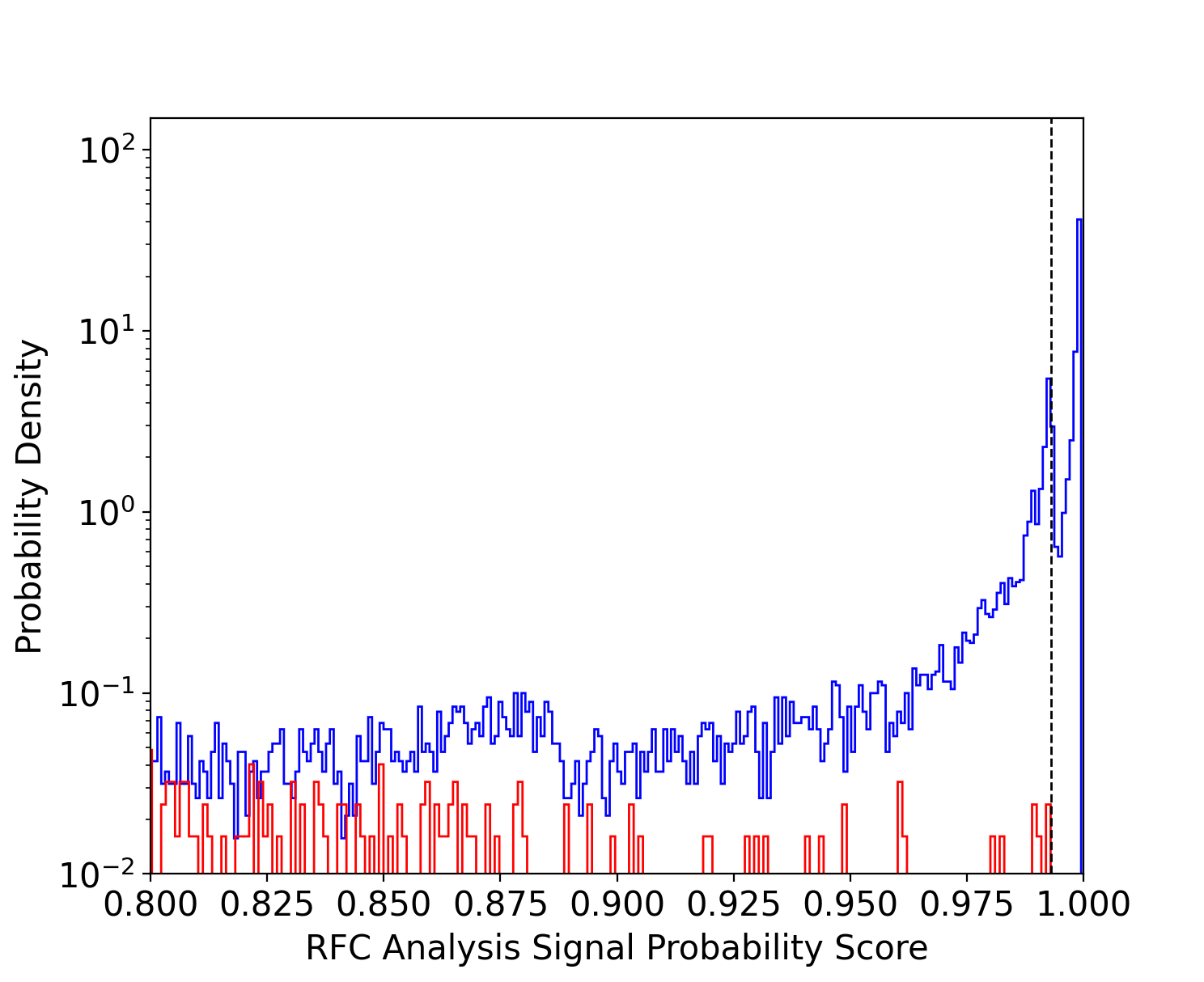}
	\caption{Distribution of signal probability scores (between 0.8 and 1.0) for the training data (blue = signal, red = background). The black dashed line shows the confidence cut. A value of 0 (1) represents a prediction of background (signal).}
	\label{fig:RFC_prob}
\end{figure}

The performance of a ML analysis generally improves with more training data. We used 80\% of our data set to train the ML model and the remaining 20\% for testing. The ML algorithm was not adjusted or tuned based on the testing set. The data selection was stratified so that the same ratio of background to signal events was maintained for both the training and testing data sets. After training, the RFC returned a signal probability score, for each training event, between 0 (most likely background) and 1 (most likely signal). A confidence cut with a value between 0 and 1 was then chosen to maximise the acceptance of signal data while removing all backgrounds in the training data. Figure \ref{fig:RFC_prob} shows the distribution of this signal probability score (zoomed into the region from 0.8 to 1.0). The vertical black dashed line in the figure shows the confidence cut.

After the confidence cut was applied, 49\% of the training signal events remained (across the whole energy range). This is the average analysis efficiency. The accuracy of the analysis model produced by the RFC was then checked using the test data set. If the model incorrectly identified a significant number of events ($\gtrsim$10 for example) from the test data background as signal, then it would not be an accurate model. Conversely, if all background events from the test data were correctly rejected but the average analysis efficiency decreased, compared to the training analysis efficiency, then the RFC model was overfitted to the training data. By fine tuning the depth and number of decision trees used by the analysis during training, the most accurate and efficient model was achieved. This was found to occur for a DT depth and number of 15 and 100, respectively, with a confidence cut of 0.993 (as shown by the black dashed line in Figure \ref{fig:RFC_prob}). This produced an average analysis efficiency (for the test data) of 47\%, which is consistent with that achieved with the training data.

In a separate study that anticipated the application of this ML approach to WIMP search data, we separated the data into three groups: the training set and test set (100 days total), and the same size WIMP search set (~55 days) as used in ref. {\cite{BATTAT201765}}. Again using an 80\%/20\% split for the training/test data (so 80 days and 20 days, respectively), we trained the ML algorithm with the training data and tested with the test data. We found an average analysis efficiency (on the test set) of 40\%. The ML model was then applied to the WIMP search data (which the ML model had never seen before). All of the events in the WIMP search data were classified as background, suggesting the absence of a WIMP signal in the dataset and validating the ML algorithm's ability to preserve sufficient background rejection for a WIMP search. 
 
\section{Detector Efficiency using ML \label{subsec:Improved_eff}}

The RFC analysis efficiency was converted into a detector efficiency by comparing the amount of neutron recoils identified by the model to that predicted by a GEANT4 \cite{AGOSTINELLI2003250} Monte Carlo simulation. The simulated results used to study the detector efficiency are those used by ref. \cite{BATTAT201765} for the same purpose. The simulation produced 9$\times{10^{8}}$ neutrons with the same energy distribution as a {$^{252}$Cf source}, originating from the $^{252}$Cf source position described in Section \ref{sec:DRIFT_data_select}. For each simulated neutron event that produced a recoil inside the DRIFT-IId gas volume, the resulting recoil type, energy and distance from the readout, $d$, was recorded. The recoil energy was converted to NIPs using known conversion factors that take into account the quenching per recoil energy and the W value of the gas. The conversion factors, up to 100 keV$_{r}$, are shown in Table \ref{tab:NIPs_convert}. 

\begin{table}[t]
	\centering
	\caption{Conversion factors for neutron induced C, F and S recoils simulated inside the DRIFT-IId detector. Table entries are in NIPs. For example, a 30 {keV$_{r}$} fluorine recoil produces 552 NIPs in the DRIFT-IId detector. Values from Ref. \cite{BATTAT201765}.}
	\label{tab:NIPs_convert}
	\begin{tabular}{c|cccccccccc}
	& \multicolumn{10}{c}{Recoil Energy [{keV$_{r}$}]} \\	
	\hline
	Species & 10 & 20 & 30 & 40 & 50 & 60 & 70 & 80 & 90 & 100 \\
	\hline
	C & 164 & 395 & 659 & 946 & 1243 & 1559 & 1877 & 2205 & 2547 & 2886 \\
	F & 140 & 332 & 552 & 792 & 1055 & 1326 & 1616 & 1911 & 2223 & 2528 \\
	S & 115 & 259 & 416 & 588 & 773 & 966 & 1167 & 1370 & 1575 & 1788 \\
	\hline
	\end{tabular}
\end{table}
		
\begin{figure}[t]
	\centering
	\includegraphics[width=\textwidth]{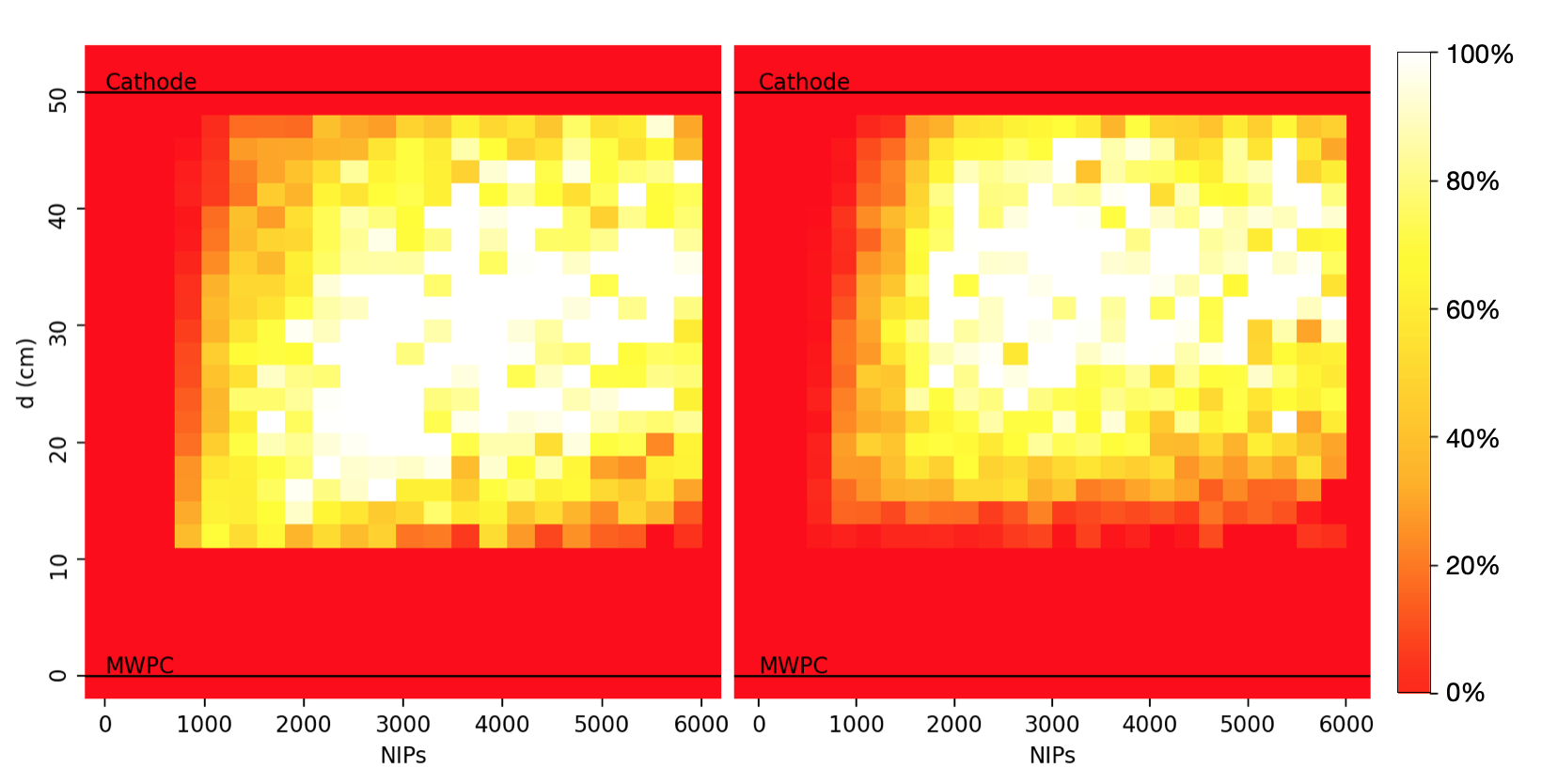}
	\caption{Efficiency maps for the previous (left) {\cite{BATTAT201765}} and this RFC analysis (right).}
	\label{fig:eff_maps}
\end{figure}

The DRIFT-IId efficiency was computed by binning in energy and $d$, with a bin width of 250 NIPs and 2 cm, respectively. The detector efficiency value in each bin is the ratio of the neutron recoil rate identified using the RFC analysis and that predicted by the simulation. The result is shown as a false colour heat map in Figure \ref{fig:eff_maps} (right), where white represents 100\% efficiency and red represents 0\% efficiency. This can be compared to the efficiency map achieved using the previous DRIFT-IId analysis \cite{BATTAT201765}, shown to the left in this figure, which uses the same false colour scale. 

\begin{figure}[t]
	\centering
	\includegraphics[scale=0.7]{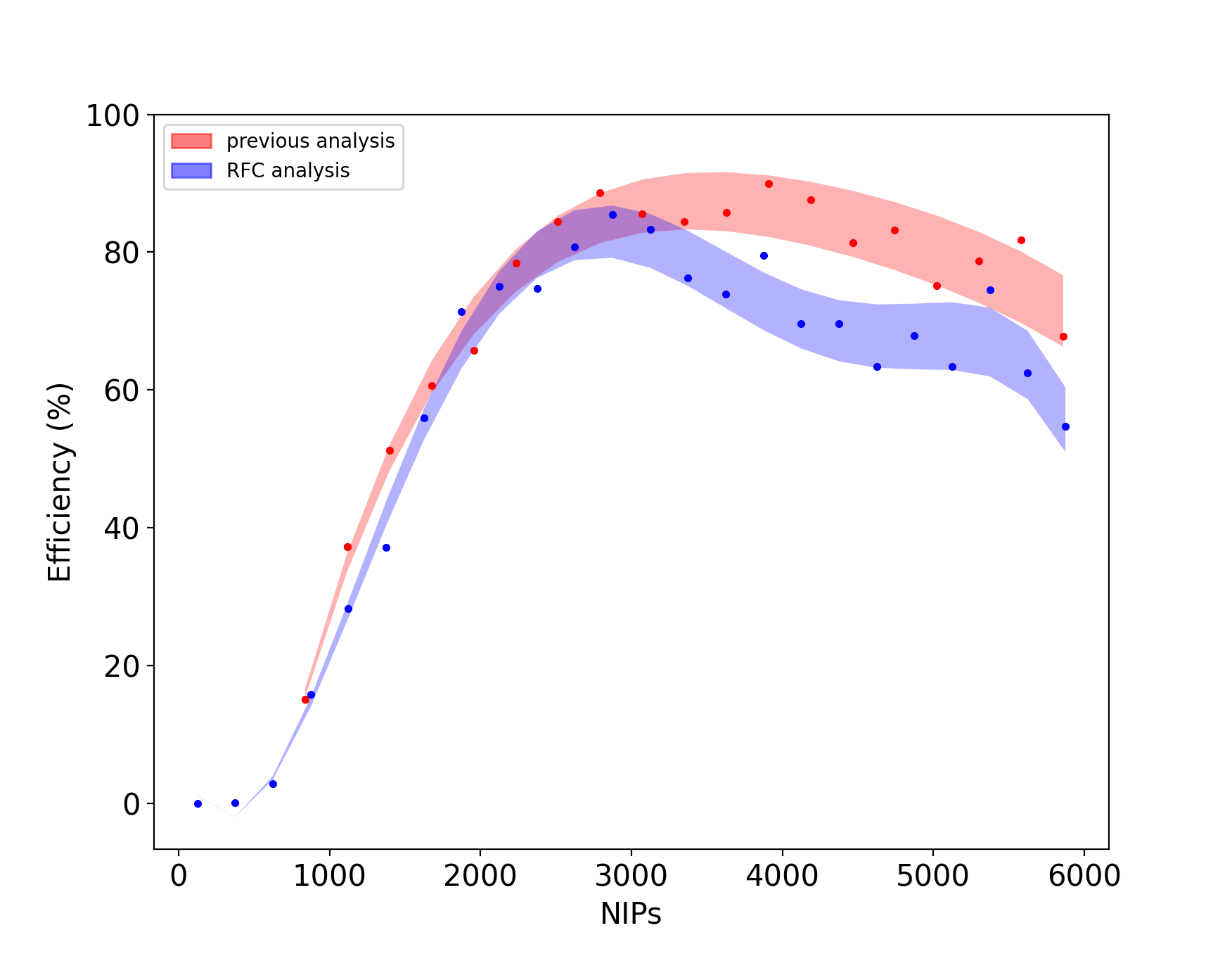}
	\caption{The detector efficiency as a function of NIPs for the previous and RFC analyses.} 
	\label{fig:avg_eff}
\end{figure}

For both analyses, Figure \ref{fig:eff_maps} shows a reduction in efficiency at high NIPs values and low $d$, and at high $d$ values and low NIPs. The former is due to high energy events producing a large main peak, which causes the peak ratio parameter cut to remove the majority of these events. The latter is due to the larger amount of diffusion experienced by charge drifting from high $d$, which dampens the signal amplitude and pushes the signal peaks below threshold. The RFC analysis efficiency, shown in Figure {\ref{fig:eff_maps}} (right) is generally comparable to that of Figure {\ref{fig:eff_maps}} (left) for the previous analysis, except that there is slightly higher efficiency evident at low NIPs and mid distance for the RFC analysis. This is illustrated more clearly in Figure {\ref{fig:avg_eff}}, which shows the detection efficiency as a function of energy, averaged over each of the NIPs bins shown in Figure {\ref{fig:eff_maps}} for the previous and RFC analysis. The upper and lower bounds of the shaded areas in Figure {\ref{fig:avg_eff}} are, respectively, 4th degree polynomial fits to the maximum and minimum Poisson standard error in efficiency for each NIPs value.

\begin{figure}[t]
	\centering
	\includegraphics[scale=0.8]{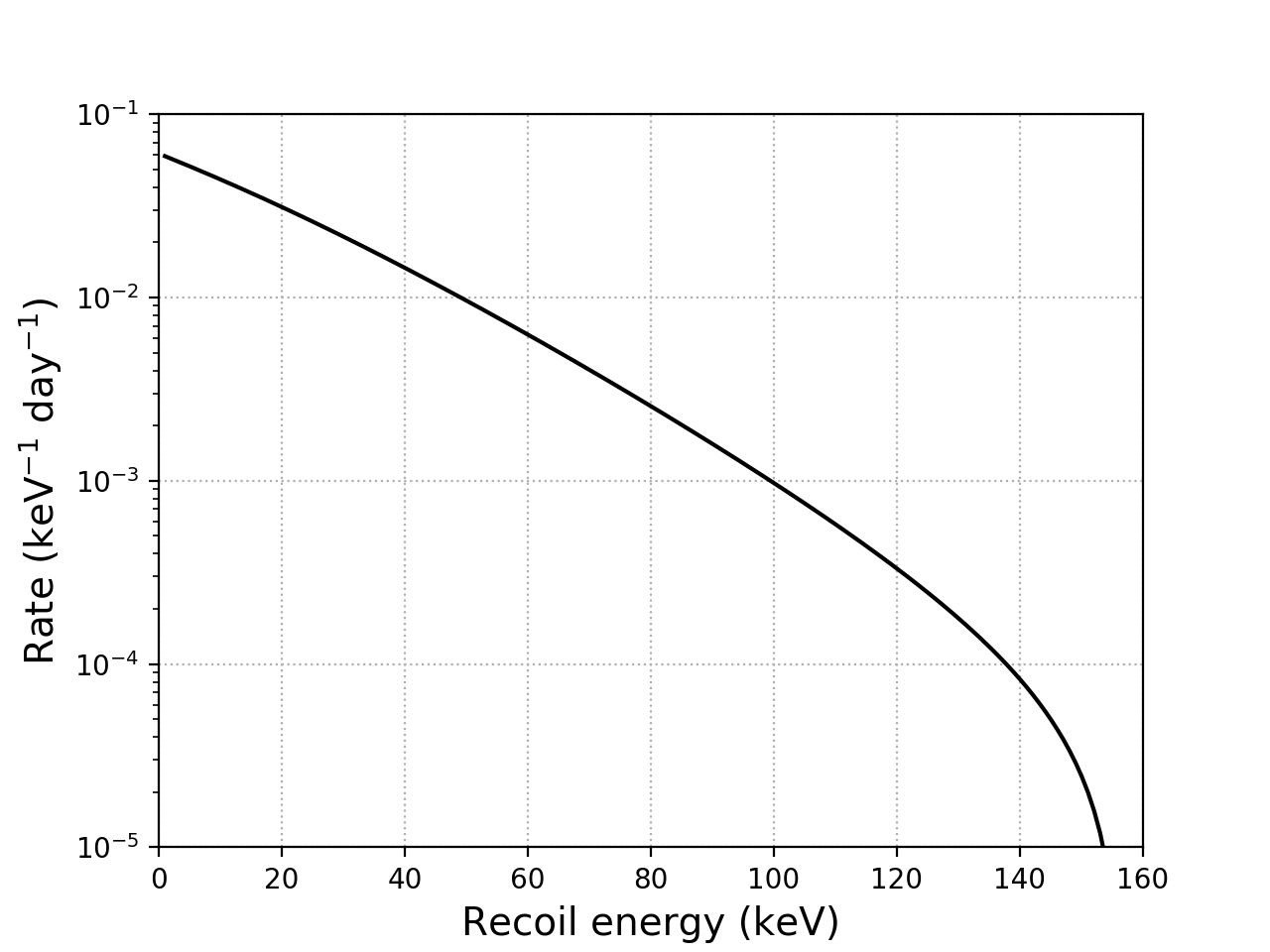}
	\caption{An example of the expected WIMP induced fluorine recoil rate as a function of recoil energy for the DRIFT-IId fiducialized fluorine mass and a WIMP mass and cross section of 100 GeV c{$^{-2}$} and 0.1 pb. Calculated using the methods described by ref. {\cite{LEWIN199687}}.}
	\label{fig:WIMP_rate}
\end{figure}

As expected, Figure \ref{fig:avg_eff} shows a drop off in efficiency at lower and higher NIPs values. This is most likely due to, respectively, the loss of minority peak information at low energy and the large main peaks that can occur at higher energies, as previously explained. Although future improvement is needed to increase the RFC efficiency at higher NIPs values in order to match the previous analysis, the result shown in Figure {\ref{fig:avg_eff}} clearly shows improved efficiency at lower NIPs values (which is more difficult to observe on the efficiency map shown in Figure {\ref{fig:eff_maps}}, right). Whereas the previous analysis has zero efficiency below 700 NIPs, the RFC analysis has an efficiency of 3\% between 500-750 NIPs and 0.4\% between 250-500 NIPs. These small efficiencies may seem trivial; however, as shown by Figure {\ref{fig:WIMP_rate}}, the WIMP rate inside the detector is expected to increase exponentially with decreasing energy, so this small increase in efficiency has a significant effect on the detector's sensitivity to WIMPs.

\section{WIMP Search Analysis}

The RFC analysis model was applied to a future hypothetical 100 days of WIMP search data. For a certain WIMP mass ($M_{W}$) and SD interaction cross section $(\sigma_{Wp}$ --where $p$ indicates that the fluorine's spin-dependancy comes from its unpaired proton), knowledge of the expected rate, $R$, of WIMP particle interactions for a given recoil energy bin, between $E_{1}$ and $E_{2}$, is given as,

\begin{equation}
	R(E_{1},E_{2}) = \int_{E_{1}}^{E_{2}}\frac{\mathrm{d}R(v_{E},v_{esc})}{\mathrm{d}E_{R}}\mathrm{d}E_{R}
	\label{eq:rate}
\end{equation}
	
The integrand is the WIMP differential rate for WIMP speeds within the interval $v_{E}$ (the average speed of the Earth, over a year, relative to the dark matter distribution) and $v_{esc}$ (the galactic escape speed). The differential rate was evaluated using the methods of ref. \cite{LEWIN199687} (eq. 3.13) and ref. \cite{TOVEY200017}, using the parameters listed in Table \ref{tab:rate_params}. 

\begin{table}[t]
	\begin{centering}
	\caption{Parameters used in the rate equation, other than $M_{W}$ and $\sigma_{Wp}$, which are unknown.}
	\label{tab:rate_params}
	\begin{tabular}{cccc}
	\toprule
	Parameter & Value & Units & ref. \\
	\midrule
	$\rho_{W}$ & 0.3 & GeV c$^{-2}$ cm$^{-3}$ & \cite{Bovy_2012}\\
	$v_{0}$ & 230 & km s$^{-1}$ & \cite{sun_speed} \\
	$v_{E}$ & 244 & km s$^{-1}$ & \cite{LEWIN199687}\\ 
	$v_{esc}$ & 600 & km s$^{-1}$ & \cite{Hattori_2018}\\
	\bottomrule
	\end{tabular}
	\par\end{centering}
\end{table}

The $v_{0}$ and $\rho_{W}$ parameters, in the above table, are the sun's orbital speed around the Galactic centre and the local dark matter density, respectively. $R$ is a function of $M_{SD}$ and the total exposure time, $t_{tot}$, which, in this case, was taken to be 100 days. $R$ is also proportional to the WIMP mass and interaction cross section such that $R \propto \sigma_{Wp}/M_{W}$.
	
\begin{figure}[t]
	\centering
	\includegraphics[scale=0.8]{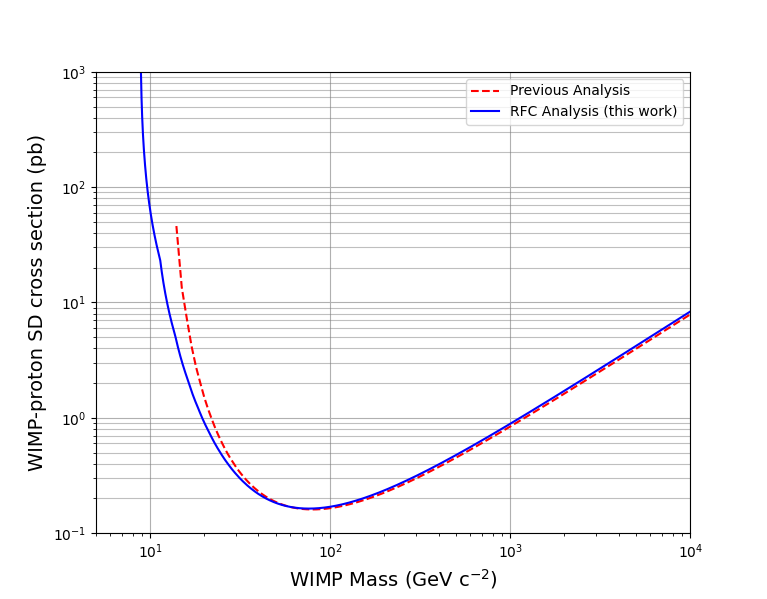}
	\caption{Projected DRIFT-IId SD WIMP exclusion limits for the previous \cite{BATTAT201765} and RFC analyses, calculated for a hypothetical 100 day exposure using the analysis efficiencies from Figure {\ref{fig:avg_eff}} and the methods and parameters described by ref. {\cite{LEWIN199687}} and ref. {\cite{TOVEY200017}}.}
	\label{fig:lim_curve}
\end{figure}

Using the methods described by Feldman and Cousins \cite{PhysRevD.57.3873}, for a particular $M_{W}$ the lowest $\sigma_{Wp}$ that can be excluded at a 90\% CL (when there is zero background leakage) is that which gives $R = 2.44$. For each $M_{W}$ between 10 and 10$^{4}$ GeV c$^{-2}$ this $\sigma_{Wp}$ was found as:

\begin{equation}
	\sigma_{Wp} = 2.44\left(\sum_{E_{min}}^{E_{max}}\epsilon(E_{R}){\frac{R(E_{1},E_{2})M_{SD}t_{tot}}{\sigma_{Wp}}}\right)^{-1}
	\label{eq:E_bin_sum}
\end{equation}

\noindent where $\epsilon(E_{R})$ is the averaged detector efficiency for the energy bin, which is given by Figure \ref{fig:avg_eff} (after converting between NIPs and recoil energy, $E_{R}$). This results in the RFC exclusion curve shown by the blue solid line in Figure \ref{fig:lim_curve}, where all $M_{W}$ and $\sigma_{Wp}$ above the curve would be excluded at 90\% CL. An exclusion curve calculated for 100 days using the previous analysis efficiency is included on this figure for comparison.  

The strongest constraint on $\sigma_{Wp}$ for the previous and RFC analyses are 0.160 pb and 0.163 pb, respectively, at WIMP masses of 80 GeV c$^{-2}$ and 76 GeV c$^{-2}$, respectively. Figure {\ref{fig:lim_curve}} shows that the previous analysis results in a slightly improved WIMP exclusion at higher WIMP masses compared to the RFC analysis, due to the previous analysis having a better efficiency at higher WIMP masses. However, the main difference in the two limit curve plots shown in Figure {\ref{fig:lim_curve}} is apparent at lower WIMP masses. Below 60 GeV {c$^{-2}$}, the RFC analysis outperforms the previous analysis, reaching as low as 9 GeV {c$^{-2}$} (compared with 14 GeV {c$^{-2}$}) and providing an order of magnitude better limit at {$M_{W}$} = 14 GeV {c$^{-2}$}. This is due to a combination of the slight increase in efficiency of the RFC analysis at lower recoil energies and the high rate of WIMP induced nuclear recoils expected at these energies (see Figure {\ref{fig:WIMP_rate}}).

\section{Conclusion}

An ML based analysis of data from the DRIFT-IId detector was presented. Using a Random Forest Classifier, we achieve enhanced detector efficiency at low recoil energy, while preserving zero background leakage. This results in an improved projected sensitivity to WIMP dark matter at masses below 60 GeV {c$^{-2}$} and a 10{$\times$} better sensitivity at 14 GeV {c$^{-2}$}. The result also indicates the feasibility of extending nuclear recoil sensitivity to WIMP masses below 10 GeV {c$^{-2}$} in an already operational direction sensitive detector for the first time. \par

This work establishes a ML analysis for directional dark matter detection using a gas based detector. Improvements can be made in future by training the analysis on a wider range of parameters and by increasing the amount of data available for training, testing, and validating the analysis algorithm.  This type of analysis will hopefully lead the way towards optimising the detection efficiency of a future large-scale, next-generation, gas-based dark matter detector, such as the one outlined by the CYGNUS collaboration \cite{Vahsen:2020pzb}, which is vital to ensure the best cost/benefit tradeoff.

For this study, previously analysed WIMP search data made up the majority of the background contribution. This real data was chosen due to the challenge of simulating the various background responses that can occur inside the detector. The signal data was emulated using a neutron source. Although this is an effecting way of inducing nuclear recoil signal inside the detector, there is an amount of background responses that can also occur during the neutron run that potentially effects the ML algorithm's ability to separate signal from background. If an accurate simulation of the various background and signal events can be produced in future, this could be used instead of or in conjunction with real data. This could potentially improve the result presented in this paper and also allow for a portion of the unused real WIMP search data to be re-analysed using a similar ML model in order to produce an actual, rather than projected, result.

\section*{Acknowledgements}

The DRIFT collaboration acknowledges the support of the Science and Technology Facilities Council (STFC), grant number: ST/N000277/1. This material is based upon work supported by the National Science Foundation under Grant Nos. 407754, 1103511, 1407773, 1506237, 1506329, 1708215, by the National Aeronautics and Space Administration under Grant No. NNX16AH49H issued through the Massachusetts Space Grant Consortium, and by the US Department of Energy under award number DE-SC0019132. The collaboration is also grateful to Israel Chemicals Ltd. (ICL) and Cleveland Potash Ltd for their support in the use of the Boulby Underground Science Facility.

\newpage

\bibliographystyle{JHEP}
\bibliography{main.bib} 

\end{document}